# Scalably manufactured high-index atomic layer-polymer hybrid metasurfaces for high-efficiency virtual reality metaoptics in the visible

Joohoon Kim[1], Junhwa Seong[1], Wonjoong Kim[2], Gun-Yeal Lee[3], Hongyoon Kim[1], Seong-Won Moon[1], Jaehyuck Jang[4], Yeseul Kim[1], Younghwan Yang[1], Dong Kyo Oh[1], Chanwoong Park[2], Hojung Choi[2], Hyeongjin Jeon[5], Kyung-Il Lee[5], Byoungho Lee[3], Heon Lee[2*], Junsuk Rho[1,4,6,7*]

[1]Department of Mechanical Engineering, Pohang University of Science and Technology (POSTECH), Pohang 37673, Republic of Korea

[2]Department of Materials Science and Engineering, Korea University, Seoul, 02841, Republic of Korea

[3]School of Electrical and Computer Engineering, Seoul National University, Seoul 08826, Republic of Korea

[4]Department of Chemical Engineering, Pohang University of Science and Technology (POSTECH), Pohang, 37673, Republic of Korea

[5]Research Institute of Industrial Science and Technology (RIST), Pohang 37673, Republic of Korea

[6]POSCO-POSTECH-RIST Convergence Research Center for Flat Optics and Metaphotonics, Pohang 37673, Republic of Korea

[7]National Institute of Nanomaterials Technology (NINT), Pohang 37673, Republic of Korea

These authors contributed equally: Joohoon Kim, Junhwa Seong, Wonjoong Kim.
*Corresponding author. E-mail: jsrho@postech.ac.kr; heonlee@korea.ac.kr

**Abstract**

Metalenses, which exhibit superior light-modulating performance with sub-micrometer-scale thicknesses, are suitable alternatives to conventional bulky refractive lenses. However, fabrication limitations, such as a high cost, low throughput, and small patterning area, hinder their mass production. Here, we demonstrate the mass production of low-cost, high-throughput, and large-aperture visible metalenses using an argon fluoride immersion scanner and wafer-scale nanoimprint lithography. Once a 12-inch master stamp is imprinted, hundreds of centimeter-scale metalenses can be fabricated. To enhance light confinement, the printed metasurface is thinly coated with a high-index film, resulting in drastic increase of conversion efficiency. As a proof of concept, a prototype of a virtual reality device with ultralow thickness is demonstrated with the fabricated metalens.

**Main**

Metalenses composed of arrays of subwavelength structures to focus light have been intensively studied to overcome the limitations of conventional optical lenses, such as chromatic aberration, the shadowing effect, existence of bulky systems, high weight, and fabrication limitations of multilevel diffractive lenses[1-7]. Achromatic metalenses have been actively demonstrated using dispersion-engineered meta-atoms[8-10]. Metalenses are light and compact because they are less than 1 μm thick. Using extremely small form factors, metalenses have been studied to replace conventional optical lenses in applications such as smartphones and virtual reality (VR) and augmented reality (AR) devices[11-13]. Despite these advantages, metalenses are challenging to commercialize owing to severe material and patterning limitations, which pose a critical bottleneck for mass production[14]. Conventional fabrication methods of visible metalenses require the thick deposition by atomic layer deposition (ALD) [2,9] or high-aspect-ratio etching[8], and high-resolution nanopatterning techniques such as electron beam lithography resulting in small pattern areas, high cost, and low throughput.

To overcome these patterning limitations, photolithography has been introduced as a method to realize large, mass-scalable metasurfaces; however, many challenges persist, such as a low patterning resolution, material limitations, and high manufacturing cost[15-18]. *i*-Line stepper lithography using a near-ultraviolet (UV, 365 nm) laser has been employed for centimeter-scale metalenses. However, the operating wavelength cannot reach the visible regime owing to its low resolution of ~400 nm[15]. Recently, krypton fluoride (KrF) stepper lithography using a mid-UV (248 nm) laser with a resolution of ~200 nm has been used for the mass production of metasurfaces, where the operating wavelength still remains in the mid- to near-infrared regimes[16,17]. An all-glass metasurface operating in the visible region (633 nm) has been demonstrated using KrF stepper lithography; however, the demonstrated metasurface has not achieved full phase modulation although the period was larger than the visible wavelength owing to the low refractive index of glass and low lithography resolutionn[18]. Recently, photolithography with shorter wavelengths such as 193 nm from an argon fluoride (ArF) laser and 13.5 nm from an extreme UV laser has been studied and used in the semiconductor industry for high-resolution patterning, but has not yet been applied in the field of nanophotonics, including metalenses, owing to its extremely high manufacturing cost.

Here, we introduce a cost-effective and high-throughput manufacturing method for visible metalenses, in which up to hundreds of centimeter-scale metalenses can be scalably fabricated once the master stamp is imprinted (**Fig. 1**). An immersion scanner with an ArF excimer laser was used to fabricate the 12-inch master stamp with a resolution of 40 nm, which was sufficiently high to fabricate the visible metalens. To overcome the high-cost limitations of ArF immersion scanners, nanoimprint lithography was used to scalably replicate the pattern of the wafer-scale master stamps at an extremely low cost. The printed metalenses were thinly coated with a high-index $TiO_2$ film to strongly confine the light, resulting in a drastic increase in the conversion efficiency from 10% to 90%. Through this simple but powerful method, we fully overcame both, patterning and material limitations, which are currently bottlenecks for the commercialization of metalenses. Our fabrication method also benefits from the mature and well-known nanoimprint lithography process, without the need for any additional complicated techniques[19]. Because the reticle, master stamp, and replica mold are reusable, the metalenses can be repeatedly replicated with high sustainability. As a proof of concept, we demonstrate a prototype of metalens-integrated VR devices that can display virtual images in each color (red, green, and blue, RGB).

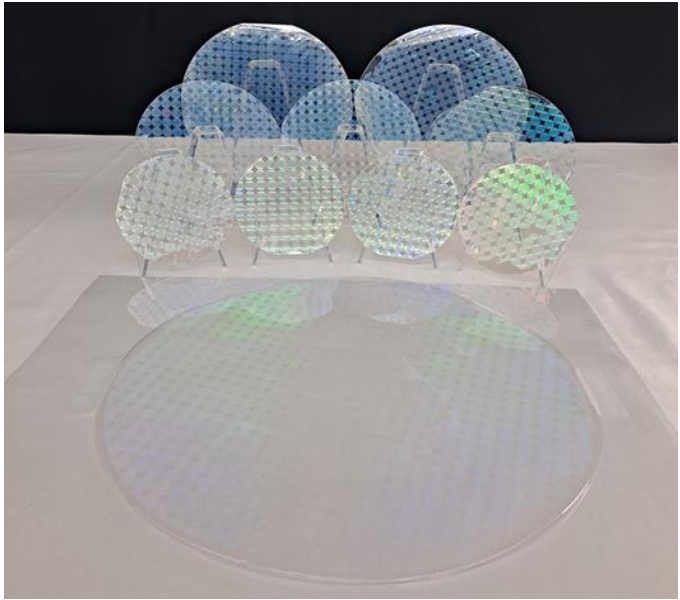

**Fig. 1. Photograph of mass-produced metalenses.** Scalable-manufactured 1-cm metalenses on 4-inch, 6-inch, 8-inch, and 12-inch wafer.

**Principle of full-phase modulation for visible metalens**

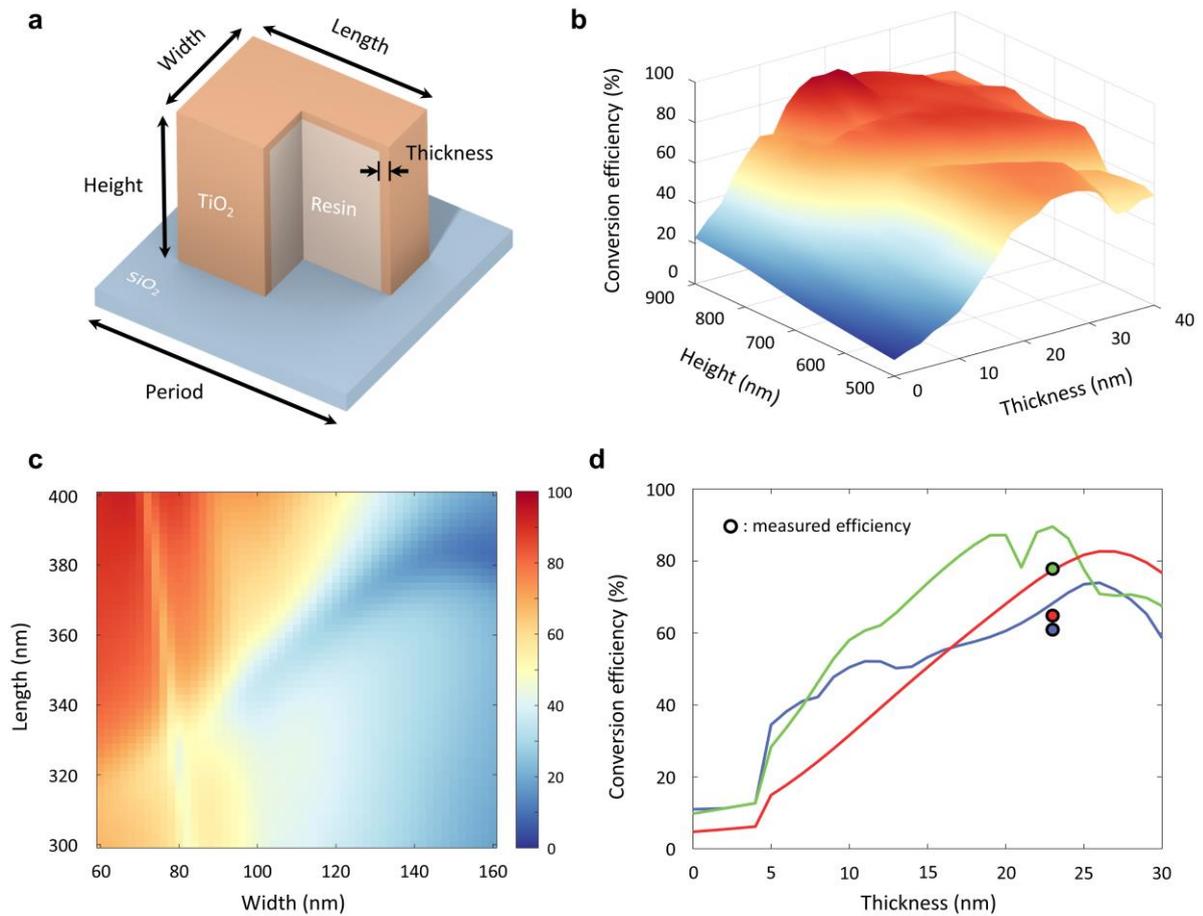

**Fig. 2. Conversion efficiencies of high-index meta-atoms. (a)** Schematic of the printable

high-index meta-atom. The meta-atom consists of the imprint resin and thinly coated $TiO_2$ film. The period was fixed at 450 nm. **(b)** Simulated maximum conversion efficiencies of meta-atoms at a wavelength of 532 nm with different heights and thicknesses. For each height and thickness, the maximum conversion efficiency was calculated from the conversion efficiency at different lengths (from 270 nm to 400 nm) and widths (from 70 nm to 200 nm). **(c)** Simulation results for meta-atoms of varying lengths and widths but equal height (900 nm) and thickness (23 nm). **(d)** Plot of conversion efficiencies simulated by varying the thickness of the $TiO_2$ film. The red, green, and blue lines represent the data at wavelengths of 635, 532, and 450 nm, respectively.

To design a high-performance metasurface, the refractive index ($n$) of the material used for the meta-atom must be sufficiently high to confine light. The main aim of this study was to use the imprinted resin itself as a meta-atom to simplify the fabrication process. However, the refractive index of the resin was approximately 1.5, which was too low to confine the light well (**Supplementary Note 1**). Therefore, a high-index film was coated on the printed resin structure to form a meta-atom, thereby increasing the effective refractive index of the metasurface (**Fig. 2a**). $TiO_2$, which has a high refractive index and low extinction coefficient in the visible region, was used as a high-index material (**Supplementary Note 2**).

In this study, a geometric phase with anisotropic meta-atoms was used to achieve full-phase modulation[2,20,21]. To design a meta-atom with a high conversion efficiency, the meta-atoms should operate as half-wave plates (**Supplementary Note 3**). We simulated the conversion efficiency of the meta-atoms by varying the width from 70 nm to 200 nm, length from 260 nm to 400 nm, height from 500 nm to 900 nm, and thickness from 0 nm to 40 nm, considering the manufacturing feasibility. As the trend of the conversion efficiencies with respect to the height and thickness was crucial, the maximum values for each height and thickness were calculated and plotted (**Fig. 2b**). We confirmed that the meta-atom with a height of 900 nm, thickness of 23 nm, length of 380 nm, and width of 70 nm had a high conversion efficiency of 89.6% at a wavelength of 532 nm (**Fig. 2c**). Moreover, the designed meta-atom had a broadband property, and exhibited high efficiencies at 532 nm and also at 450 nm (68.2%) and 635 nm (79.7%) (**Supplementary Note 4**). As expected, for all three wavelengths (450, 532, and 635 nm), the maximum conversion efficiency drastically increased as the thickness increased, which verified that the coating of the high-index material increased the effective refractive index (**Fig. 2d**). **Supplementary Note 5** shows that the high-index coating ensured the strong confinement of light, whereas the pure resin meta-atom exhibited weak confinement. Finally, we experimentally confirmed that the designed meta-atom has conversion efficiency of 77.8% (532 nm), 64.8% (635 nm), 60.9% (450 nm) by deflecting the conversed light (**Supplementary Note 6**).

**Fabrication method for mass-production of visible metalenses**

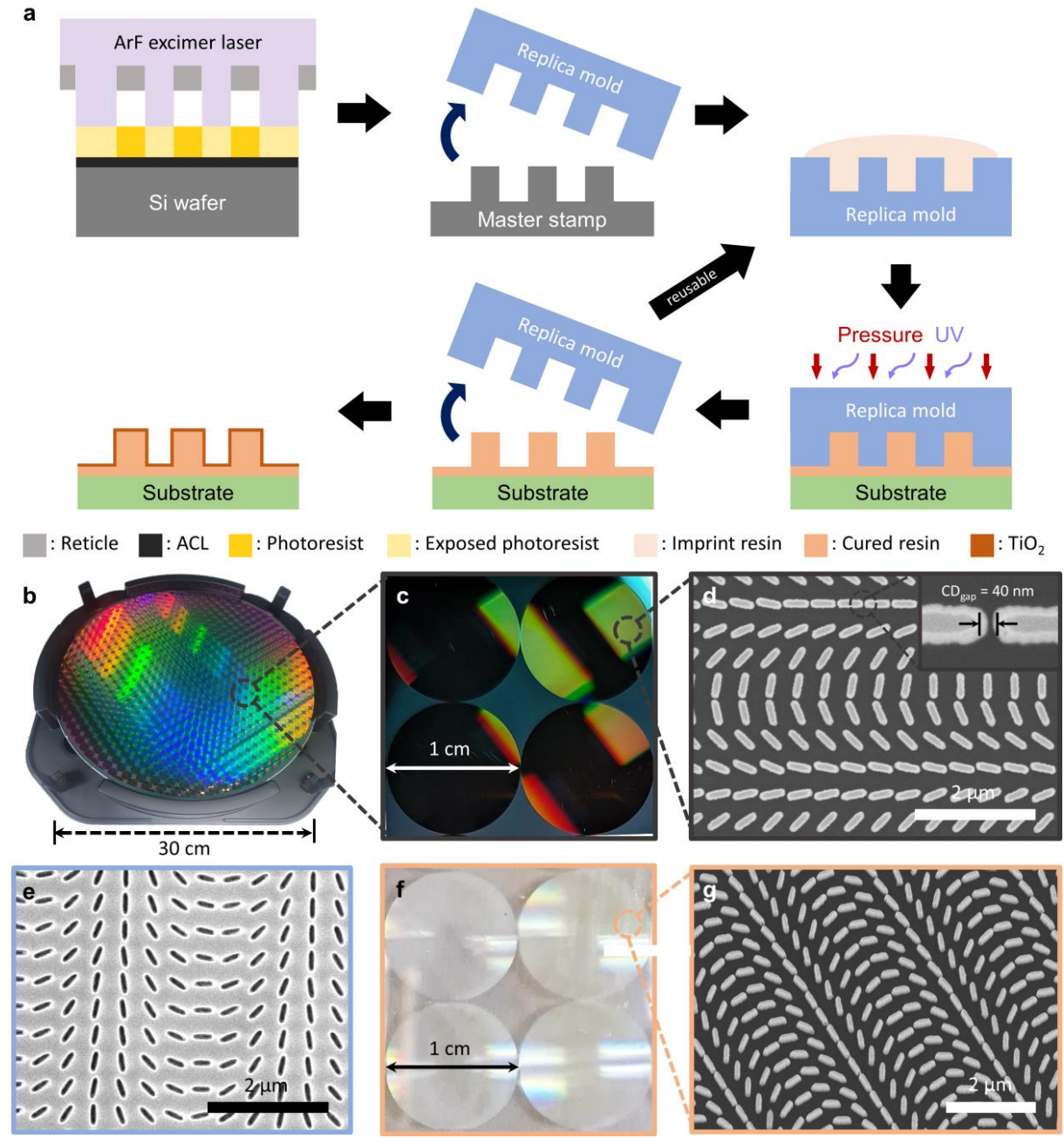

**Fig. 3. Mass production of high-efficiency visible metalenses using ArF immersion scanner** Fabrication process for mass production of high-index visible metalenses using ArF immersion scanner. **(a)** Fabrication schematic. **(b,c)** Photograph of the fabricated 12-inch

master stamp. **(d)** Scanning electron microscope (SEM) image of the fabricated master stamp. (insets) Zoom-in SEM image of gap between meta-atoms. **(e)** SEM image of the fabricated replica mold. **(f)** Photograph of the replicated metalenses with a diameter of 1 cm. **(g)** 5° tilted SEM image of the replicated metalens with high-index coating. All scale bars: 2 μm.

A schematic of the proposed fabrication process is shown in **Fig. 3a**. First, a 4X reticle for the ArF immersion scanner is written using electron-beam lithography. Using an ArF immersion scanner with a resolution of 40 nm, 669 dies of 1 cm metalens are continuously exposed on a 12-inch Si wafer. In sequence, the pattern of the exposed photoresist is transferred to the Si wafer via a deposition and etching process to form a master mold (**Fig. 3b–d, Supplementary Note 7**). Next, a hard polydimethylsiloxane (*h*-PDMS) solution is coated and solidified on the master mold to replicate the metasurface with a high resolution of 70 nm owing to its high modulus (~10 N/mm$^2$)[22]. Then, PDMS, which acts as a flexible buffer layer, is coated onto the *h*-PDMS layer to form the replica mold (**Fig. 3e, Supplementary Note 8**). The imprint resin is dropped onto the replica mold and covered with a glass substrate. MINS-311RM is used as a UV-curable imprint resin because of its low surface energy, low shrinkage (<5%), and relatively high refractive index. The imprinted resin is fully cured under pressure and UV light irradiation. After the replica mold is released, resin nanostructures are created on the glass substrate. Because the imprint resin has low surface energy, the resin nanostructures can be easily detached from the replica mold without any surface treatment. A thin layer of TiO$_2$ is then coated onto the printed resin nanostructures via ALD (**Fig. 3f,g**). In this method, metalenses can be scalably imprinted from 4-inch to 12-inch glass wafer (**Fig. 1**).

Our fabrication method features large-area and high-resolution patterning of high-index dielectric nanostructures at low cost and high throughput. Notably, the use of an ArF immersion scanner, nanoimprint lithography instrument, and ALD apparatus, enable the mass production of metalenses. First, an ArF immersion scanner with a resolution of 40 nm is used to fabricate a 12-inch master stamp that can replicate 669 metalenses once imprinted. High resolution not only enables width critical dimension (CD) of 75 nm but also gap CD of 40 nm (**Fig. 3d**). Both length and width are fabricated larger than the target size in consideration of the contraction of the resin. Second, nanoimprint lithography, which can repeatedly replicate master stamp patterns at an extremely low cost, is employed to circumvent the high fabrication cost of ArF photolithography. Moreover, nanoimprint lithography has high throughput, and requires only 15 min to replicate the pattern from the replica mold to the substrate. Optimizing the intensity of the UV lamp may shorten the replication time. Furthermore, our fabrication method is highly sustainable because the reticle, master stamp, and replica mold are reusable. Finally, ALD is used to increase the effective refractive index of meta-atoms. All the three fabrication steps are also compatible with mature and well-known semiconductor fabrication methods, without the need for additional facilities..

**Characterization of the fabricated metalens**

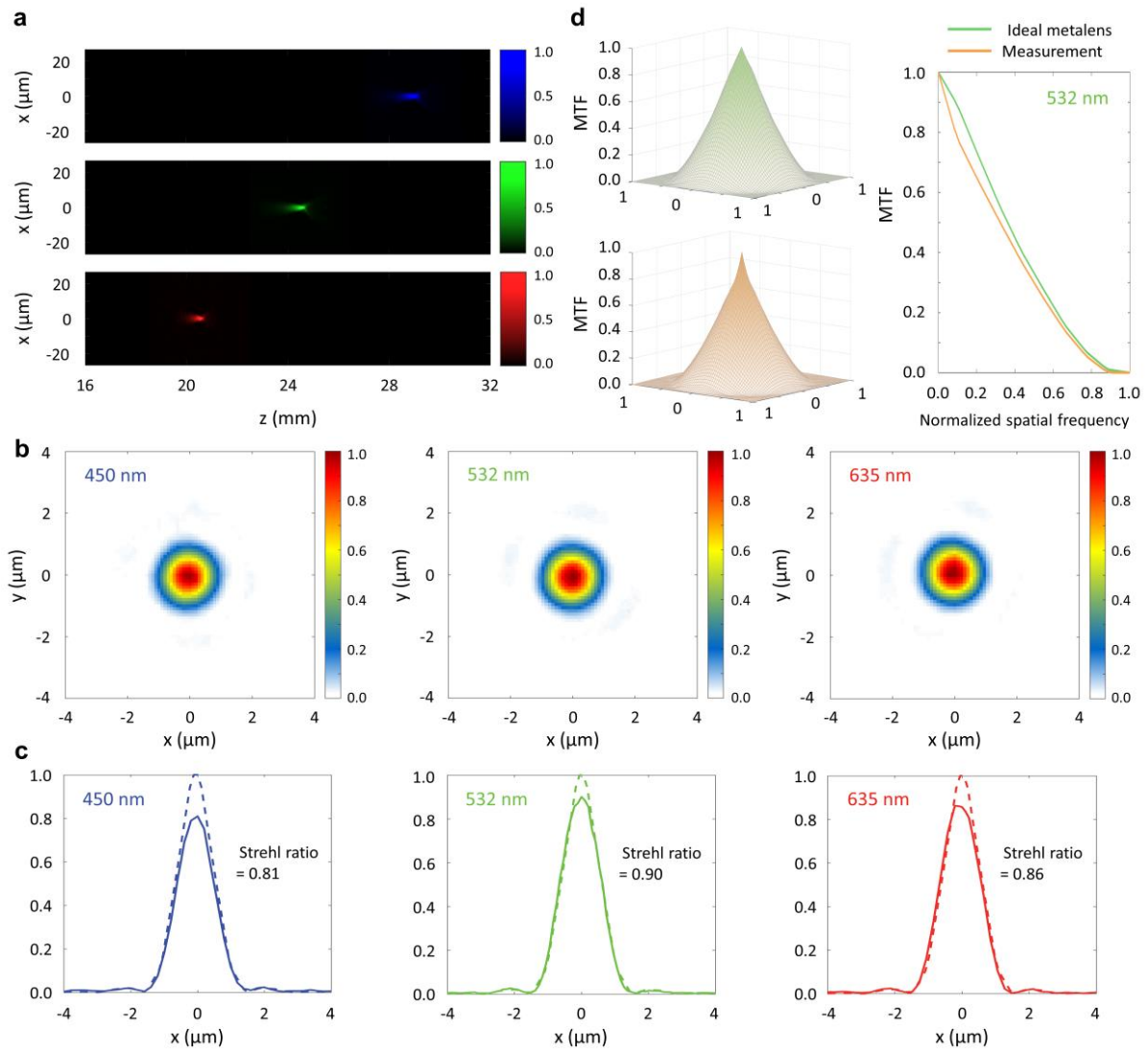

**Fig. 4. Optical characterization of the metalens. (a)** Measured intensity profiles of the focal spot in the X-Z plane at different wavelengths (450 nm, 532 nm, and 635 nm). **(b)** Measured normalized intensity distributions of the focal planes. **(c)** Corresponding focusing intensity line profiles of (b). **(d)** Modulation transfer function (MTF) comparison between the ideal imaging system and the metalens at 532 nm.

The 1-cm diameter metalens was designed to follow the ideal phase distribution ($\varphi_{ideal}$) and focus the cross-polarized component of the transmitted light, given by

$$\varphi_{ideal} = -\frac{2\pi}{\lambda}\left(\sqrt{x^2 + y^2 + f^2} - f\right) \qquad (2)$$

where $\lambda$ = 532 nm is the design wavelength, $x$ and $y$ are the spatial coordinates of each meta-atom on the metalens, $f$ = 2.45 cm is the design focal length with a numerical aperture (NA) = 0.2. To fulfill the ideal phase distribution, meta-atoms were rotated by $\theta = \frac{\varphi_{ideal}}{2}$ at each position (**Eq. 1**). The Rayleigh-Sommerfeld diffraction formula[23] can be used to simulate the focusing characteristics with discretized phases (**Supplementary Note 9**). As a result of chromatic aberration, the focal spot of two different wavelengths shifted 2.90 cm at 450 nm and 2.05 cm at 635 nm, corresponding to NA = 0.17 and 0.24, respectively. By applying the above principles, a 1-cm diameter metalens with an NA of 0.2 was designed at 532 nm.

To evaluate the performance of the fabricated metalens, the focusing properties were characterized using a customized measurement setup (**Supplementary Note 10**). The intensity profiles of the focal spot for different wavelengths were measured on the X-Z plane (**Fig. 4a**). The focal length and corresponding NA value at the RGB wavelengths were consistent with the simulated results, including a negligible variation. The diffraction efficiency, which can be directly connected to conversion efficiency, was also characterized as 20.9% at 450 nm, 34.4% at 532 nm, and 23.0% at 635 nm[24]. Using the measured diffraction efficiency, we calculated a veiling glare of 0.37 at 450 nm, 0.51 at 532 nm, and 0.46 at 635 nm; the overall performance metrics were 0.24, 0.38, and 0.26, respectively[25]. The normalized intensity distributions on the focal plane were demonstrated with the corresponding Strehl ratio (SR) (**Fig. 4b, c**). Each SR was near unity at different wavelengths, i.e., 0.81 at 450 nm, 0.90 at 532 nm, and 0.86 at 635 nm, indicating that they can act as an ideal imaging system.

Furthermore, to analyze the imaging capability of the metalens, we calculated the modulation transfer functions (MTFs) using simulated point spread functions (PSF). The MTF, which is defined as the amplitude of the Fourier transformation of the PSF, offers metalens performance in terms of resolution and contrast. Thus, the MTF of the metalens can be expressed as[26]

$$MTF \equiv \left| \frac{\iint I(x,y)\exp[-i2\pi(f_x x + f_y y)]\,dxdy}{\iint I(x,y)\,dxdy} \right| \qquad (3)$$

where $I(x,y)$ indicates the PSF, and $f_x$ and $f_y$ represent the spatial frequencies along the *x*- and *y*-axes, respectively. The MTF curves of the ideal imaging system and the fabricated metalens were compared at the design wavelength (**Fig. 4d**). The cutoff frequency of $2NA/\lambda$, denoting that the NA value and wavelength limit the spatial resolution, was used for normalizing the spatial frequency. The MTF curves of the measured metalens were analogous to the ideal diffraction-limited lenses at 532 nm and the other two wavelengths (**Supplementary Note 11**). Therefore, we confirmed that mass-manufactured metalenses can follow the ideal lens equation with near-ideal focusing.

**Metalens-driven virtual reality device**

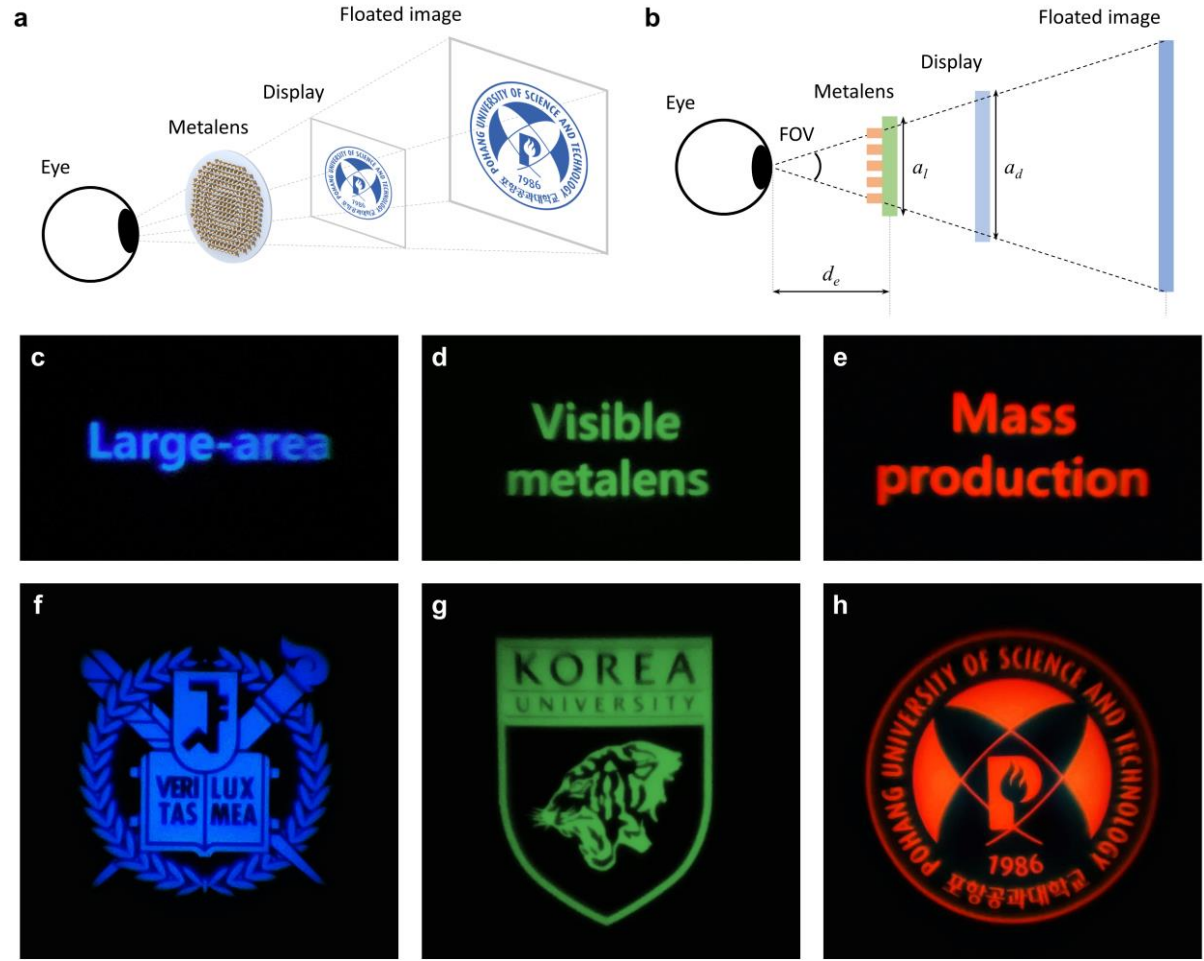

**Fig. 5. Metalens-integrated virtual information imaging.** (**a**) Schematic illustration of the near-eye display system for VR with mass-produced metalens. (**b**) Cross-section view of the system with detailed parameters. The VR imaging results using a passive display showing (**c-e**) letters and (**f-h**) logos in three colors: (**c,f**) blue; (**d,g**) green; (**e,h**) red. The images shown

here were captured by a cell phone. FOV: field of view.

As a proof of concept, mass-manufactured metalenses were applied to demonstrate VR devices. Recently, VR devices have garnered significant attention as next-generation displays owing to their promising ability to provide vivid three-dimensional visual experiences[27,28]. VR devices primarily require compact, lightweight, and large-aperture lenses for a comfortable user experience. We demonstrate a near-eye VR display system with lightweight and ultrathin metalenses. **Fig. 5a** shows a schematic illustration of the proposed system using a mass-produced metalens. Unlike conventional VR devices, in which virtual images are reflected or diffracted onto the user's eye, in this device, the metalens are located in front of the eye as a transmissive eyepiece to reduce the light path. As shown in **Fig. 5b**, the maximum field of view (FOV) is determined by the eye relief distance ($d_e$) and lens aperture ($a_l$): FOV = $2\tan^{-1}(a_l/2d_e)$. Thus, in a transmission-type eyepiece, the FOV can be increased without limitations by enlarging the lens aperture. Based on the proposed display system with a high-performance metalens, a prototype of the VR device was demonstrated (**Fig. 5c-h**). Virtual images were captured using a system comprising a display panel, circular polarizers, and metalens.

The main advantage of metalenses for VR devices is their lightweight and compact form factor. In conventional VR devices, the distance of the focal length between the display panel and the imaging system enlarge the device form factor, resulting in an increase in the device volume while keeping most of it empty. Our metalenses which work as transmissive eyepieces can address this limitation and achieve a compact form factor. Although our compact system proves that our mass-produced metalens can be successfully applied to VR imaging, the performance of our metalens can be further improved. The focal length can be reduced by increasing the NA of the lens. High-NA metalenses enable a wider FOV than conventional see-through near-eye displays. Finally, achromatic metalenses enable the full-color imaging of virtual information. Notably, high-NA and achromatic metalenses for high-performance VR/AR devices have already been introduced, and their design method is well-known[12,13]. We believe that high-performance VR/AR devices with compact form factors can be commercialized by combining our fabrication method with the aforementioned design methods.

**Conclusions**

In the last two decades, metalenses that replace diffractive optics and provide additional advantages have been intensively studied; however, metalenses have not been commercialized because of difficulties in meeting commercial manufacturing capabilities. In summary, we fully addressed this long-standing challenge by introducing a simple but powerful method for the mass production of centimeter-scale metalenses with low cost and high throughput. Our method is based on an ArF immersion scanner, nanoimprint lithography, and ALD, which can benefit from the mature and well-known semiconductor fabrication methods. An ArF immersion scanner with a resolution of 40 nm was used to fabricate a 12-inch master stamp for centimeter-scale visible metalenses. Using nanoimprint lithography, patterns of the master stamp could be replicated on a substrate at an extremely low cost and high throughput. Finally, ALD was used for the thin coating of high-index materials to increase the effective refractive index. By thinly coating printed metalenses with the high-index material, conversion efficiencies were drastically increased for all three wavelengths: 450 nm (68.2%), 532 nm (89.6%), and 635 (79.7%). We confirmed that mass-manufactured metalenses have diffraction efficiencies of 20.9% at 450 nm, 34.4%% at 532 nm, and 23.0% at 635 nm. In this method, we mass-produced hundreds of centimeter-scale metalenses, even in a laboratory environment, not at mass production facilities. As a proof of concept, a prototype of a metalens-integrated VR device was demonstrated. The demonstrated VR devices can display virtual information for the RGB colors with an extremely low form factor and are lightweight. Our method can also be applied to the mass production of various metasurfaces[29] such as bound states in the continuum[30], biosensors[31,32], color printing[33-35], and hologram[36-38]. We believe that this work will enable the introduction of metasurfaces into the industry.

**Methods**

*Refractive indices measurement*: The refractive indices of the imprint resin and high-index $TiO_2$ film were measured using ellipsometry. The measured amplitude ratio and phase difference between the *s* and *p* components were fitted using the Cauchy dispersion model. Model coefficients for the imprint resin and $TiO_2$ film were determined with a mean squared error of 7.340 and 11.324, respectively.

*Numerical simulation*: All results in this work were calculated using the FEM-based commercial software (COMSOL Multiphysics 6.0). All simulations were performed within the *xyz* space with periodic boundary conditions.

*Master stamp fabrication*: A molybdenum silicide on a silica substrate was used as the reticle. The reticle for metalenses was transferred onto a positive photoresist (FEP series, Fujifilm Ltd.) using high-power e-beam lithography (JBX Series, JEOL Ltd.). The exposed patterns were developed in tetramethylammonium (TMAH). A 69-nm thick molybdenum silicide layer was deposited using an e-beam evaporator. A Si substrate was used as the cluster mold. Patterns of the reticle were transferred onto positive-tone photoresists using ArF lithography (XT-1900Gi, ASML). The exposed patterns were developed in TMAH.

*Nanoimprint lithography*: For the demolding of *h*-PDMS, the cluster mold is coated with a liquid phase self-assembly monolayer (SAM) solution. The SAM solution is prepared by mixing hexane and (heptadecafluoro-1,1,2,2-tetra-hydodecyl) trichlorosilane (H5060.1, JSI silicone). *h*-PDMS prepared by 3.4 g of vinylmethyl siloxane (VDT-731, Gelest), 18 μL of platinum catalyst (SIP6831.2, Gelest), 0.1 g of modulator (2,4,6,8 – tetramethyl-2,4,6,8-tetravinylcyclotetrasiloxane, Sigma-aldrich), 2g of toluene, and 1g of curing agent (HMS-301, Gelest) is spin-coated in two steps of 500rpm for 10sec and 3000rpm for 50sec, then baked 80 °C for 2h. PDMS mixture is prepared by mixing PDMS (Sylgard 184 A, Dow corning) and curing agent (Sylgard 184 B, Dow corning) and degassed in the vacuum chamber for 30min to remove air bubbles. The mixture is poured onto the cluster mold and cured at 80 °C for 3h to peel off the *h*-PDMS layer form the cluster mold. UV curable resin (MINS-311RM, Minuta Tech) is poured on the fabricated stamp and is placed on a glass substrate with UV resin side down. The sample is pressed with 2 bar in UV irradiation (IMDE04-A01, Jungang automatic technology) to cure the UV resin and transfer the metalens template. Finally, UV resin metalens template is created on the glass substrate after demolding the fabricated stamp at room temperature.

*TiO$_2$ deposition condition*: On top of the fabricated resin nanostructures, a plasma-enhanced ALD TiO$_2$ thin film was deposited using a thermal and plasma time-divided ALD station (NexusBe, Nexus Plaminar™ Series). Tetrakis(dimethylamido) titanium(IV) (TDMAT, (Ti(NMe$_2$)$_4$)) was used as the Ti precursor, and oxygen plasma (350 W, 1 Torr) was utilized as the reactant. The deposition process consisted of 0.5 s TDMAT pulses, a 15 s argon purge, 5 s oxygen plasma exposure, and a 15 s argon purge at 80 °C in the ALD chamber. The growth rate per cycle was 1.0 Å/cycle.

**Data availability**

The data supporting the findings of this study are available from the corresponding author upon reasonable request.

**Author contributions**

J. R., H. L., and J. K. proposed the idea and conceived the experiment; J. K. and J. S. contributed to writing the manuscript; J. K., H. K., S.-W. M., and Y. K. performed the theoretical and numerical simulations; W. K., C. P., and H. C. contributed to the fabrication of the master mold and nanoimprinting; J. S. performed the atomic layer deposition of $TiO_2$; J. S. and J. K. performed the experimental characterization and data analysis; G.-Y. L., J. J., and B. L. demonstrated the VR/AR devices; all the authors contributed to writing the final manuscript; J. R. and H. L. guided the entire project.

**Acknowledgements**

POSCO-POSTECH-RIST Convergence Research Center program funded by POSCO

National Research Foundation (NRF) grants (##) funded by the Ministry of Science and ICT of the Korean government.

National Research Foundation (NRF) grants (NRF-2019K1A47A02113032) funded by the Ministry of Science and ICT of the Korean government.

Technology Innovation Program (20016234, Nano-molding materials with ultra-High refractive index for meta-lens) funded by the Ministry of Trade, Industry & Energy (MOTIE, Korea)

**References**

1  Chen, X. *et al.* Dual-polarity plasmonic metalens for visible light. *Nature Communications* **3**, 1198, doi:10.1038/ncomms2207 (2012).
2  Khorasaninejad, M. *et al.* Metalenses at visible wavelengths: Diffraction-limited focusing and subwavelength resolution imaging. *Science* **352**, 1190–1194, doi:10.1126/science.aaf6644 (2016).
3  Chen, K. *et al.* A Reconfigurable active huygens' metalens. *Advanced Materials* **29**, 1606422, doi:10.1002/adma.201606422 (2017).

4	Schlickriede, C. *et al.* Imaging through nonlinear metalens using second harmonic generation. *Advanced Materials* **30**, 1703843, doi:10.1002/adma.201703843 (2018).
5	Yoon, G., Kim, K., Huh, D., Lee, H. & Rho, J. Single-step manufacturing of hierarchical dielectric metalens in the visible. *Nature Communications* **11**, 2268, doi:10.1038/s41467-020-16136-5 (2020).
6	Zhou, Y., Zheng, H., Kravchenko, I. I. & Valentine, J. Flat optics for image differentiation. *Nature Photonics* **14**, 316–323, doi:10.1038/s41566-020-0591-3 (2020).
7	She, A., Zhang, S., Shian, S., Clarke, D. R., & Capasso, F. Adaptive metalenses with simultaneous electrical control of focal length, astigmatism, and shift. *Science Advances* **4**, eaap9957, doi:10.1126/sciadv.aap9957 (2018).
8	Wang, S. *et al.* A broadband achromatic metalens in the visible. *Nature Nanotechnology* **13**, 227–232, doi:10.1038/s41565-017-0052-4 (2018).
9	Chen, W. T. *et al.* A broadband achromatic metalens for focusing and imaging in the visible. *Nature Nanotechnology* **13**, 220–226, doi:10.1038/s41565-017-0034-6 (2018).
10	Lin, R. J. *et al.* Achromatic metalens array for full-colour light-field imaging. *Nature Nanotechnology* **14**, 227–231, doi:10.1038/s41565-018-0347-0 (2019).
11	Lee, G.-Y. *et al.* Metasurface eyepiece for augmented reality. *Nature Communications* **9**, 4562, doi:10.1038/s41467-018-07011-5 (2018).
12	Li, Z. *et al.* Meta-optics achieves RGB-achromatic focusing for virtual reality. *Science Advances* **7**, eabe4458, doi:10.1126/sciadv.abe4458 (2021).
13	Li, Z. *et al.* Inverse design enables large-scale high-performance meta-optics reshaping virtual reality. *Nature Communications* **13**, 2409, doi:10.1038/s41467-022-29973-3 (2022).
14	Engelberg, J. & Levy, U. The advantages of metalenses over diffractive lenses. *Nature Communications* **11**, 1991, doi:10.1038/s41467-020-15972-9 (2020).
15	She, A., Zhang, S., Shian, S., Clarke, D. R., & Capasso, F. Large area metalenses: design, characterization, and mass manufacturing. *Opt. Express* **26**, 1573–1585, doi:10.1364/OE.26.001573 (2018).
16	Tao, J. *et al.* Mass-manufactured beam-steering metasurfaces for high-speed full-duplex optical wireless broadcasting communications. *Advanced Materials* **n/a**, 2106080, doi:10.1002/adma.202106080.
17	Leitis, A., Tseng, M. L., John-Herpin, A., Kivshar, Y. S., & Altug, H. Wafer-scale functional metasurfaces for mid-infrared photonics and biosensing. *Advanced Materials* **33**, 2102232, doi:10.1002/adma.202102232 (2021).
18	Park, J.-S. *et al.* All-glass, large metalens at visible wavelength using deep-ultraviolet projection lithography. *Nano Letters* **19**, 8673–8682, doi:10.1021/acs.nanolett.9b03333 (2019).
19	Kumar, K. *et al.* Printing colour at the optical diffraction limit. *Nature Nanotechnology* **7**, 557–561, doi:10.1038/nnano.2012.128 (2012).
20	Zheng, G. *et al.* Metasurface holograms reaching 80% efficiency. *Nature Nanotechnology* **10**, 308–312, doi:10.1038/nnano.2015.2 (2015).
21	Lin, D., Fan, P., Hasman, E., & Brongersma, M. L. Dielectric gradient metasurface optical elements. *Science* **345**, 298–302, doi:10.1126/science.1253213 (2014).
22	Schmid, H. & Michel, B. Siloxane polymers for high-resolution, high-accuracy soft lithography. *Macromolecules* **33**, 3042-3049, doi:10.1021/ma982034l (2000).
23	Shen, F. & Wang, A. Fast-Fourier-transform based numerical integration method for the Rayleigh-Sommerfeld diffraction formula. *Appl. Opt.* **45**, 1102–1110, doi:10.1364/AO.45.001102 (2006).

24      Engelberg, J. & Levy, U. Standardizing flat lens characterization. *Nature Photonics* **16**, 171–173, doi:10.1038/s41566-022-00963-7 (2022).
25      Engelberg, J. & Levy, U. A True Assessment of Flat Lenses for Broadband Imaging Applications. *arXiv preprint arXiv:2107.12830* (2021).
26      Goodman, J. W. *Introduction to Fourier Optics*. (McGraw-Hill, 1996).
27      Song, J.-H., van de Groep, J., Kim, S. J., & Brongersma, M. L. Non-local metasurfaces for spectrally decoupled wavefront manipulation and eye tracking. *Nature Nanotechnology* **16**, 1224–1230, doi:10.1038/s41565-021-00967-4 (2021).
28      Faisal, A. Computer science: Visionary of virtual reality. *Nature* **551**, 298–299, doi:10.1038/551298a (2017).
29      Ni, J. *et al.* Multidimensional phase singularities in nanophotonics. *Science* **374**, eabj0039, doi:10.1126/science.abj0039 (2021).
30      Koshelev, K., Lepeshov, S., Liu, M., Bogdanov, A., & Kivshar, Y. Asymmetric metasurfaces with high-$Q$ resonances governed by bound states in the continuum. *Physical Review Letters* **121**, 193903, doi:10.1103/PhysRevLett.121.193903 (2018).
31      Tittl, A. *et al.* Imaging-based molecular barcoding with pixelated dielectric metasurfaces. *Science* **360**, 1105–1109, doi:10.1126/science.aas9768 (2018).
32      Yesilkoy, F. *et al.* Ultrasensitive hyperspectral imaging and biodetection enabled by dielectric metasurfaces. *Nature Photonics* **13**, 390–396, doi:10.1038/s41566-019-0394-6 (2019).
33      Ruan, Q. *et al.* Reconfiguring colors of single relief structures by directional stretching. *Advanced Materials* **34**, 2108128, doi:10.1002/adma.202108128 (2022).
34      Liu, H. *et al.* Tunable resonator-upconverted emission (TRUE) color printing and applications in optical security. *Advanced Materials* **31**, 1807900, doi:10.1002/adma.201807900 (2019).
35      Jang, J. *et al.* Spectral modulation through the hybridization of Mie-scatterers and quasi-guided mode resonances: Realizing full and gradients of structural color. *ACS Nano* **14**, 15317–15326, doi:10.1021/acsnano.0c05656 (2020).
36      Fang, X., Ren, H., & Gu, M. Orbital angular momentum holography for high-security encryption. *Nature Photonics* **14**, 102–108, doi:10.1038/s41566-019-0560-x (2020).
37      Bao, Y., Wen, L., Chen, Q., Qiu, C.-W., & Li, B. Toward the capacity limit of 2D planar Jones matrix with a single-layer metasurface. *Science Advances* **7**, eabh0365, doi:10.1126/sciadv.abh0365 (2021).
38      Song, Q., Odeh, M., Zúñiga-Pérez, J., Kanté, B., & Genevet, P. Plasmonic topological metasurface by encircling an exceptional point. *Science* **373**, 1133–1137, doi:10.1126/science.abj3179 (2021).

# Supplementary Information

# Scalably manufactured high-index atomic layer-polymer hybrid metasurfaces for high-efficiency virtual reality metaoptics in the visible

Joohoon Kim[1], Junhwa Seong [1], Wonjoong Kim[2], Gun-Yeal Lee[3], Hongyoon Kim[1], Seong-Won Moon[1], Jaehyuck Jang[4], Yeseul Kim[1], Younghwan Yang[1], Dong Kyo Oh[1], Chanwoong Park[2], Hojung Choi[2], Hyeongjin Jeon[5], Kyung-Il Lee[5], Byoungho Lee[3], Heon Lee[2*], Junsuk Rho[1,4,6,7*]

[1]Department of Mechanical Engineering, Pohang University of Science and Technology (POSTECH), Pohang 37673, Republic of Korea

[2]Department of Materials Science and Engineering, Korea University, Seoul, 02841, Republic of Korea

[3]School of Electrical and Computer Engineering, Seoul National University, Seoul 08826, Republic of Korea

[4]Department of Chemical Engineering, Pohang University of Science and Technology (POSTECH), Pohang, 37673, Republic of Korea

[5]Research Institute of Industrial Science and Technology (RIST), Pohang 37673, Republic of Korea

[6]POSCO-POSTECH-RIST Convergence Research Center for Flat Optics and Metaphotonics, Pohang 37673, Republic of Korea

[7]National Institute of Nanomaterials Technology (NINT), Pohang 37673, Republic of Korea

These authors contributed equally: Joohoon Kim, Junhwa Seong, Wonjoong Kim.

*Corresponding author. E-mail: jsrho@postech.ac.kr; heonlee@korea.ac.kr

**Supplementary Note 1. Measured refractive index and extinction coefficient of the imprint resin.**

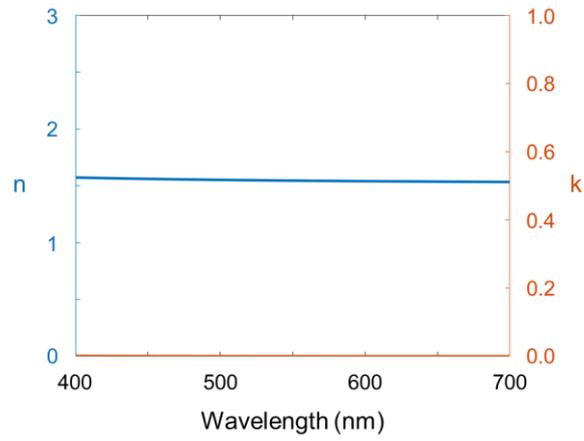

**Supplementary Figure 1.** Measured refractive index and extinction coefficient of the imprint resin in the visible regime.

**Supplementary Note 2. Measured refractive index and extinction coefficient of the TiO$_2$ film.**

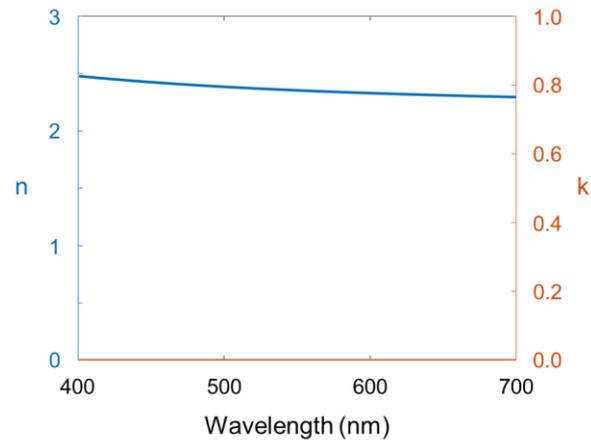

**Supplementary Figure 2.** Measured refractive index and extinction coefficient of the TiO$_2$ film in the visible regime.

**Supplementary Note 3. Design principle of rotated anisotropic meta-atoms.**

In this work, anisotropic meta-atoms are used to achieve full-phase modulation. The optical responses of $\theta$ rotated anisotropic meta-atoms can be analyzed with Jones calculus. Anisotropic meta-atoms can be represented by Jones matrix **J**, given by

$$\mathbf{J} = \begin{bmatrix} t_{xx} & 0 \\ 0 & t_{yy} \end{bmatrix} \tag{S1}$$

where $t_{xx}$ and $t_{yy}$ represent the complex transmission coefficients. The first and second subscripts represent the polarization of the incident and transmitted light, respectively. The key concept of the full-phase modulation using anisotropic meta-atoms is rotating the meta-atom. The Jones matrix A for meta-atoms rotated by an angle $\theta$ can be represented using a rotation matrix $R(\theta)$, as follows

$$\mathbf{A} = \mathbf{R}(-\theta)\mathbf{J}\mathbf{R}(\theta) = \begin{bmatrix} \cos\theta & \sin\theta \\ -\sin\theta & \cos\theta \end{bmatrix} \begin{bmatrix} t_{xx} & 0 \\ 0 & t_{yy} \end{bmatrix} \begin{bmatrix} \cos\theta & -\sin\theta \\ \sin\theta & \cos\theta \end{bmatrix} \tag{S2}$$

For the incident right-handed circularly polarized (RCP) wave, the Jones vector of the outgoing wave $\mathbf{E}_T$ can be calculated as follows

$$\begin{aligned}\mathbf{E}_T = \mathbf{A}\frac{1}{\sqrt{2}}\begin{bmatrix}1\\-i\end{bmatrix} &= \begin{bmatrix} t_{xx}\cos^2\theta + t_{yy}\sin^2\theta + i(t_{yy} - t_{xx})\sin\theta\cos\theta \\ (t_{yy} - t_{xx})\sin\theta\cos\theta + i(t_{yy}\cos^2\theta + t_{xx}\sin^2\theta) \end{bmatrix} \\ &= \frac{t_{xx}+t_{yy}}{2}\begin{bmatrix}1\\-i\end{bmatrix} + \frac{t_{xx}-t_{yy}}{2}e^{-i2\theta}\begin{bmatrix}1\\i\end{bmatrix}\end{aligned} \tag{S3}$$

The first and second subscripts represent the polarization of the incident and transmitted light, respectively. The outgoing wave consists of two orthogonal components which are a co-polarized component (RCP) and a cross-polarized component (LCP). Notably, the cross-polarized component experiences a phase delay of -2$\theta$, enabling the full-phase modulation by rotating the meta-atom from 0 to π. Therefore, the amplitude of the cross-polarized component is defined as the conversion efficiency, which is directly connected to the efficiency of the metalens. Due to its wavelength independence, designed anisotropic meta-atoms have the advantage of the broadband property, opening the possibility of full-color metalenses.

**Supplementary Note 4. Broadband property of the designed meta-atom.**

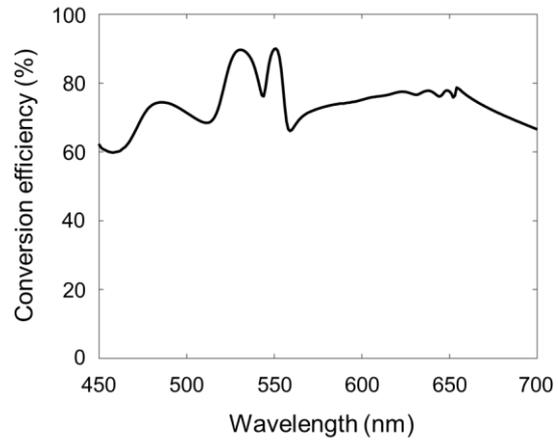

**Supplementary Figure 3.** Plot of conversion efficiency with respect to wavelength.

**Supplementary Note 5. Simulated electric field profiles of meta-atoms.**

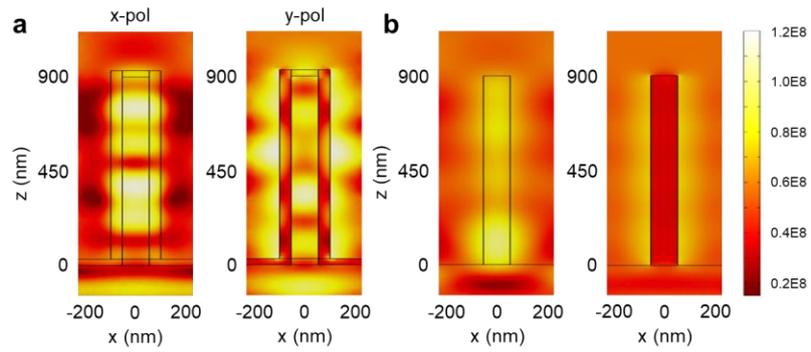

**Supplementary Figure 4.** Simulated electric field profiles of meta-atoms consisting of **(a)** TiO$_2$ coated resin and **(b)** pure resin. The real part of the *x* component (left) and *y* component (right) at the design wavelength of 532 nm.

**Supplementary Note 6. Calculation of the conversion efficiency.**

| Wavelength | Steering intensity | Incident intensity | Conversion efficiency |
|---|---|---|---|
| 450 nm | 2.60 µW | 4.27 µW | 60.9% |
| 532 nm | 3.79 µW | 4.87 µW | 77.8% |
| 635 nm | 9.72 µW | 15.01 µW | 64.8% |

**Supplementary Table 1.** Beam steering intensity measurement.

The conversion efficiency of the designed meta-atom is experimentally measured with the metasurface designed for beam steering. The conversion efficiency is calculated as the ratio of the measured intensity from the deflected beam and the total incident intensity.

**Supplementary Note 7. Fabrication method of the master stamp.**

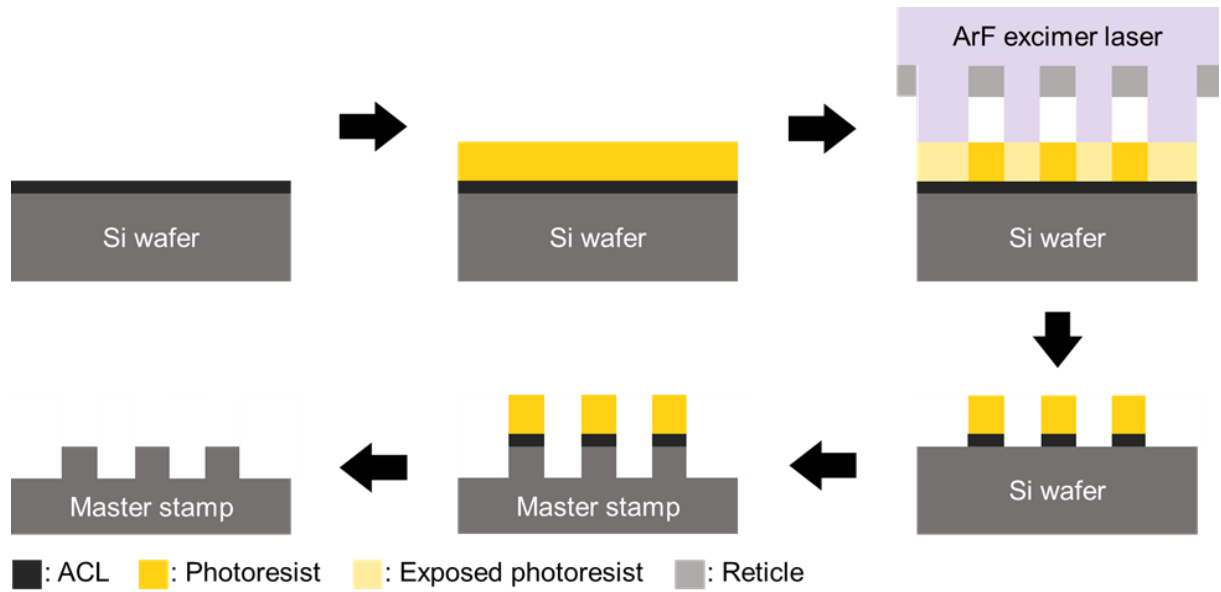

**Supplementary Figure 5.** The detailed fabrication schematic of the master stamp.

A brief schematic of the master stamp fabrication process is provided in Fig. S5a. First, the amorphous carbon layer (ACL) is deposited on the 12-inch silicon (Si) wafer to compensate weak etch rate of photoresist. Then, the photoresist (PR) is spin-coated on the ACL deposited Si wafer. 600 dies of 1 cm metalens are exposed on the wafer via argon fluoride (ArF) excimer laser. The exposed wafer is developed and carbon layer is etched. The pattern is transferred to the Si wafer by reactive ion etching (RIE) using the PR and ACL as the etch mask. Finally, the remaining etch mask is removed by the PR stripper and $O_2$ plasma.

**Supplementary Note 8. Images of the fabricated replica mold.**

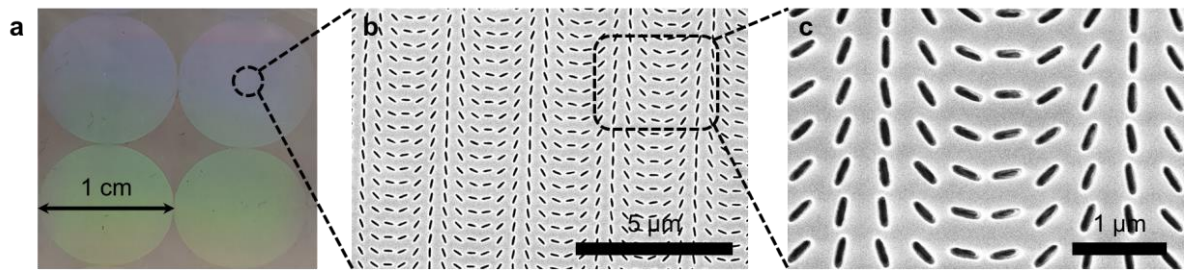

**Supplementary Figure 6.** Images of the fabricated replica molds. (a) Photography of the fabricated replica molds. (b,c) Top-view SEM images of the fabricated replica mold. Scale bar in (b) and (c) represent 5 μm and 1 μm, respectively.

**Supplementary Note 9. Simulated intensity profiles of the focal spot in the X-Z plane using Rayleigh Sommerfeld diffraction formula.**

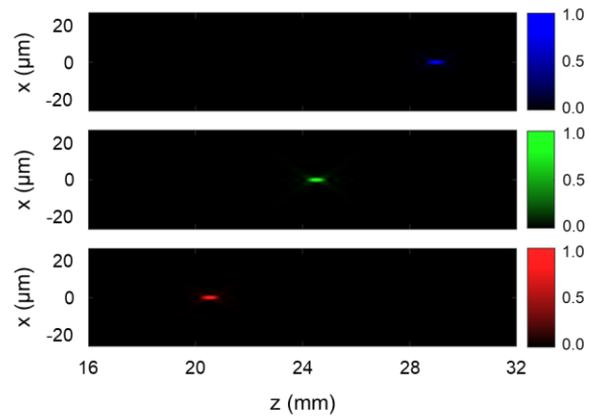

**Supplementary Figure 7.** Simulated intensity profiles of the focal spot in the X-Z plane at 450, 532, and 635 nm.

**Supplementary Note 10. Customized measurement setup to characterize the focusing properties of the metalens.**

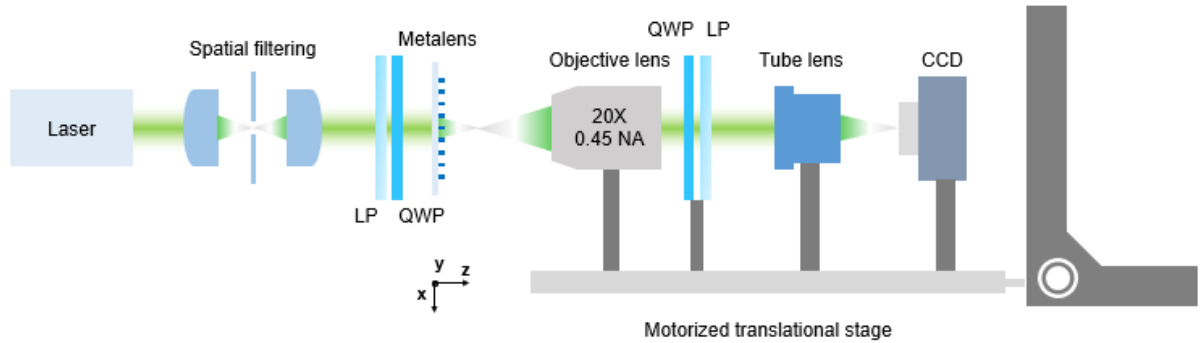

**Supplementary Figure 8.** The focal spot of the metalens is characterized using an optimized optical setup. A laser beam passes through to the spatial filtering setup (Thorlabs, AC127-019-A, Thorlabs, P10CB) for generating a clean Gaussian beam. The collimated laser beam is incident to a linear polarizer (Thorlabs, LPVISE050-A) and quarter waveplate (Thorlabs AQWP05M-600) to produce the left-handed circularly polarized (LCP) beam. The normally incident LCP beam is focused by the metalens, then the diverging beam is collected by a commercial objective lens (Olympus, LUCPLFLN20X). To single out the conjugate right-handed circularly polarized (RCP) beam, an analyzer is placed in between the objective lens and tube lens (Thorlabs, TTL180-A). Therefore, the CCD camera (Lumenera, INFINITY2-1RC) receives only a cross-polarization component corresponding to a PSF. The intensity profiles of the focal spot are measured around at each focal point. Further, we have calculated the veiling glare as on the ratio of the diffraction efficiency to the transmission and overall performance metric (OPM) as $OPM = \frac{\eta}{\sqrt{T}} \times SR$, where $\eta$ is the measured diffraction efficiency, T is the overall transmission, and SR is the one-dimensional Strehl ratio.

**Supplementary Note 11. MTF comparison between ideal imaging system and the metalens at two different wavelengths.**

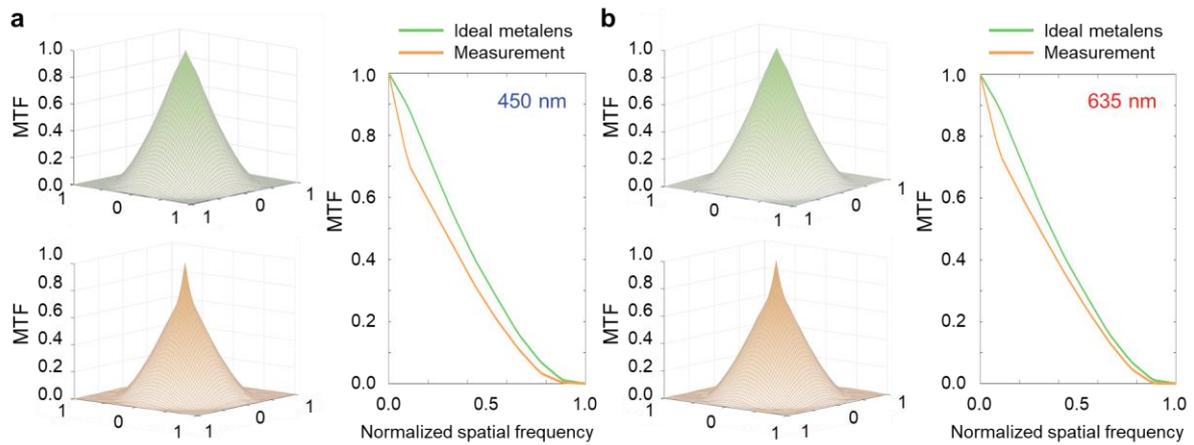

**Supplementary Figure 9.** MTF comparison between the ideal imaging system and the fabricated metalens (a) at 450 nm and (b) at 635 nm. The SR is calculated as the ratio of volume under the two-dimensional MTF curve to the volume under the diffraction-limited two-dimensional MTF curve.